\documentclass[prd,showkeys,showpacs,superscriptaddress]{revtex4}
\usepackage{amsmath}
\usepackage{amssymb}
\usepackage{graphicx}
\usepackage{float}
\usepackage{xcolor}
\usepackage{physics}
\usepackage{ulem}

\allowdisplaybreaks

\begin{document}

\title{Charged 	spinning fermionic configurations and a mass gap
}

\author{
Vladimir Dzhunushaliev
}
\email{v.dzhunushaliev@gmail.com}
\affiliation{
Department of Theoretical and Nuclear Physics,  Al-Farabi Kazakh National University, Almaty 050040, Kazakhstan
}
\affiliation{
Institute of Nuclear Physics, Almaty 050032, Kazakhstan
}
\affiliation{
Academician J.~Jeenbaev Institute of Physics of the NAS of the Kyrgyz Republic, 265 a, Chui Street, Bishkek 720071, Kyrgyzstan
}
\affiliation{
Laboratory for Theoretical Cosmology, International Centre of Gravity and Cosmos,
Tomsk State University of Control Systems and Radioelectronics (TUSUR),
Tomsk 634050, Russia
}

\author{Vladimir Folomeev}
\email{vfolomeev@mail.ru}
\affiliation{
Institute of Nuclear Physics, Almaty 050032, Kazakhstan
}
\affiliation{
Academician J.~Jeenbaev Institute of Physics of the NAS of the Kyrgyz Republic, 265 a, Chui Street, Bishkek 720071, Kyrgyzstan
}
\affiliation{
Laboratory for Theoretical Cosmology, International Centre of Gravity and Cosmos,
Tomsk State University of Control Systems and Radioelectronics (TUSUR),
Tomsk 634050, Russia
}

\begin{abstract}
We consider a self-consistent axially symmetric system supported by a classical nonlinear spinor field minimally coupled to electric and magnetic Maxwell fields. 
The presence of the nonlinearity of the spinor field ensures the existence of a minimum positive energy of the system (a mass gap), of a minimum charge (a charge gap), 
and of a minimum magnetic moment. In turn, the presence of the electric charge results in qualitative changes in the behavior of physical characteristics of the 
systems under consideration as compared with the case of an electrically neutral spinor field. It is shown that, with a suitable choice of free system parameters, 
there exists a regular finite-energy particlelike solution describing a localized spinning object whose physical parameters correspond to the main characteristics 
of an electron/positron (including the spin equal to $1/2$), but with the characteristic size comparable to the corresponding Compton wavelength.
Also, we show that four local Dirac equations are equivalent to two nonlocal equations.
\end{abstract}


\date{\today}

\maketitle

\section{Introduction}

Nonlinear equations describing various physical systems have been the object of numerous investigations in different aspects. 
The bulk of such studies have been mainly focused on a consideration of the nonlinear Shr\"{o}dinger and Klein-Gordon equations involving different potentials. 
The first of these equations permits one to describe various phenomena and processes within condensed matter physics, nonlinear optics,
atomic and mathematical physics. In turn, the Klein-Gordon equation is widely used in modeling various particlelike objects in condensed matter and mathematical physics, 
including strongly gravitating systems.

Much less attention was paid to investigations of the nonlinear Dirac equation. Such an equation was initially introduced by D.~Ivanenko~\cite{Ivanenko:1938}. 
Subsequently, it was analyzed in the works~\cite{Finkelstein:1951zz,Finkelstein:1956}, where possible forms of nonlinear terms were suggested. 
Following these ideas, W.~Heisenberg  tried to employ this equation as a fundamental equation suitable for describing the properties of an electron~\cite{Heisenberg}. 
Later, one of forms of the nonlinear Dirac equation was employed for an approximate description of the properties of hadrons (this approach is called 
the Nambu-Jona-Lasinio model~\cite{Nambu:1961}; for a review, see Ref.~\cite{Volkov:2006}).
In turn, bearing in mind that the systems with a nonlinear spinor field contain a mass gap~\cite{Finkelstein:1951zz,Finkelstein:1956,Soler:1970xp,Ranada:1973hna}, 
in Ref.~\cite{Ranada:1974hx}, the authors tried to describe extended particles (hadrons) possessing the smallest possible energy. 
On the other hand, in the case of fermions with zero bare mass, Ref.~\cite{Gross:1974jv} suggests a toy model of quark confinement in quantum chromodynamics. 
In addition, the nonlinear Dirac equations may be used as effective theories in various fields of atomic, nuclear, particle, and gravitational physics~\cite{Alvarez:1981yh,Ionescu:1987jr,Esteban:1995,Esteban:2002,Zecca:2002dq,Bronnikov:2004uu,Bronnikov:2009na,Adanhounme:2012cm,Saha:2016cbu,
Dzhunushaliev:2018jhj,Dzhunushaliev:2019kiy,Dzhunushaliev:2019uft,Bronnikov:2019nqa}.

In quantum chromodynamics, there is a well-known problem to prove the existence of a minimum value of the mass, i.e., of a mass gap,  
in non-Abelian quantum Yang-Mills theory. This problem is very nontrivial and has not been solved yet. One of possible approaches towards solving this problem might be 
a consideration of simpler problems when quantum systems are replaced by some approximate classical systems. In this case, if one could show that for such 
classical configurations a mass gap might occur, this could be treated as a possible indication for the existence of the mass gap in quantum systems. 
As shown in our previous investigations~\cite{Dzhunushaliev:2019ham,Dzhunushaliev:2020qwf,Dzhunushaliev:2021apa,Dzhunushaliev:2022pkz}, 
in the systems supported by classical non-Abelian fields coupled to nonlinear spinor fields, there is the possibility of obtaining a mass gap. 
From this point of view, such classical systems  may be thought of as approximately describing realistic quantum systems.

The systems considered by us earlier in Refs.~\cite{Dzhunushaliev:2019ham,Dzhunushaliev:2020qwf,Dzhunushaliev:2021apa,Dzhunushaliev:2022pkz} are spherically symmetric. 
Obviously, in the more general case the spherical symmetry can already be violated, for example, because of the presence of magnetic Maxwell and/or color fields. 
Correspondingly,  this requires a generalization of the above models. As a first step in this direction, one can consider a simplified situation where the system 
with a nonlinear spinor field contains only Abelian fields. Consistent with this, the present paper studies a system supported by a classical nonlinear spinor field 
minimally coupled to Maxwell electric and magnetic (dipole) fields. Due to the presence of the dipole magnetic field, the system is inevitably axisymmetric, 
and therefore a consideration will be carried out in a general form without any simplifying assumptions as to the smallness of the electromagnetic fields, 
as was done earlier, for example, in Refs.~\cite{Ranada:1973hna,Ranada:1974hx}.
This will enable us to study the cases where the contribution to the energy-momentum tensor coming from the electromagnetic fields can be comparable 
to that of the spinor field. It will be shown that such a contribution results in qualitative changes in the physical characteristics of the systems under consideration.

Notice here that in the present paper we consider a system supported by a classical spinor field. Following
Ref.~\cite{ArmendarizPicon:2003qk}, such a field is meant to be a set of four complex-valued spacetime functions transforming according to the 
spinor representation of the Lorentz group. In turn, realistic spin-1/2 particles must evidently be described by quantum spinor fields, and it is furthermore 
believed that there is no classical limit for quantum spinor fields. However, classical spinors can be thought of as arising from some effective description 
of more complex quantum systems (for arguments in favor of the possibility of the existence of classical spinors, see Ref.~\cite{ArmendarizPicon:2003qk}).

\section{The model}

We consider localized configurations consisting of a spinor field $\psi$ minimally coupled to Maxwell fields. 
The corresponding total Lagrangian for such a system can be represented in the form
(we use natural units with $c=\hbar=1$ throughout)
\begin{equation}
\label{Lagr_gen}
	L_{\text{tot}} = \frac{\imath }{2} \left(
			\bar \psi \gamma^\mu \psi_{; \mu} -
			\bar \psi_{; \mu} \gamma^\mu \psi
		\right) - m  \bar \psi \psi - F(S)-\frac{1}{4} F_{\mu\nu}F^{\mu\nu} ,
\end{equation}
where $m$ is  a bare mass of the fermion,  $F(S)$ is in general an arbitrary nonlinear term with the invariant $S$ (see below), 
and the electromagnetic field tensor $F_{\mu\nu} = \partial_{ \mu} A_\nu - \partial_\nu A_\mu$. The semicolon denotes the covariant derivative defined as
$
\psi_{; \mu} =  [\partial_{ \mu} +1/8\, \omega_{a b \mu}\left( \gamma^a  \gamma^b- \gamma^b  \gamma^a\right)+\imath\, Q A_\mu]\psi
$ with $\gamma^a$ being the Dirac matrices in flat space; the term $\imath\, Q A_\mu\psi$ describes the interaction between 
the spinor and Maxwell fields with the coupling constant $Q$. In turn, the Dirac matrices in curvilinear coordinates, $\gamma^\mu = e_a^{\phantom{a} \mu} \gamma^a$, 
are obtained using the tetrad 
$ e_a^{\phantom{a} \mu}$, and $\omega_{a b \mu}$ is the spin connection [for its definition, see Ref.~\cite{Lawrie2002}, Eq.~(7.135)]. 
In the above expressions, $\mu,\nu=0,1,2,3$ are spacetime indices and $a,b=0,1,2,3$ are tetrad indices. In what follows, we use the Weyl representation of the Dirac matrices,
$$
\gamma^0 =
     \begin{pmatrix}
        0   &   1 \\
        1   &   0
    \end{pmatrix},\quad
\gamma^k =
     \begin{pmatrix}
        0   &   \sigma^k \\
        -\sigma^k   &   0
    \end{pmatrix},
$$ 
where $k=1,2,3$ and $\sigma^k$ are the Pauli matrices.

Varying the action with the Lagrangian~\eqref{Lagr_gen} with respect to the spinor field and to the vector potential $A_\mu$, we derive the corresponding Dirac and Maxwell field equations
\begin{eqnarray}
	\imath  \gamma^\mu \psi_{;\mu} - m \psi - \frac{\partial F}{\partial\bar\psi}&=& 0,
\label{feqs-20}\\
	\frac{1}{\sqrt{-g}}\frac{\partial}{\partial x^\nu}\left(\sqrt{-g}F^{\mu\nu}\right)&=&-Q\bar\psi\gamma^\mu\psi.
\label{feqs-22}
\end{eqnarray}
In the present paper, we take the following simplest self-interaction term
$$
	F(S) = - \frac{\lambda}{2} \left(\bar\psi\psi\right)^2 ,
$$
where $\lambda$ is some free nonlinearity parameter.
Classical spinor fields with such a nonlinearity have been considered, for instance,
in Refs.~\cite{Finkelstein:1951zz,Finkelstein:1956,Soler:1970xp,Ranada:1973hna,Ranada:1974hx,Dzhunushaliev:2018jhj,Dzhunushaliev:2019kiy,Dzhunushaliev:2019uft}.

From the Lagrangian~\eqref{Lagr_gen}, one can also obtain the corresponding energy-momentum tensor of the system under consideration (already in a symmetric form)
\begin{equation}
\label{EM}
	T_{\mu}^\nu = \frac{\imath }{4}g^{\nu\rho}
	\left[
		\bar\psi \gamma_{\mu} \psi_{;\rho} 
		+ \bar\psi\gamma_\rho\psi_{;\mu}
	- \bar\psi_{;\mu}\gamma_{\rho }\psi 
	- \bar\psi_{;\rho}\gamma_\mu\psi
	\right] 
	- \delta_\mu^\nu L_{\text{sp}}
	- F^{\nu\rho} F_{\mu\rho} 
	+ \frac{1}{4} \delta_\mu^\nu F_{\alpha\beta} F^{\alpha\beta}.
\end{equation}
Taking into account the Dirac equation \eqref{feqs-20} and the corresponding adjoint equation for $\bar\psi$, the Lagrangian for the spinor field appearing in Eq.~\eqref{EM} becomes
$$
	L_{\text{sp}} = - F(S) + \frac{1}{2} \left(
		\bar\psi\frac{\partial F}{\partial\bar\psi} +
		\frac{\partial F}{\partial\psi}\psi
	\right).
$$

In the present paper, we take the stationary {\it Ansatz} for the spinor field in the form similar to that of Ref.~\cite{Herdeiro:2019mbz},
\begin{equation}
	\psi = e^{\imath \left(M\varphi-\Omega t\right)} 
		\begin{pmatrix}
			\psi_1 \\ \psi_2 \\ \psi_2^*\\\psi_1^*
		\end{pmatrix},
\label{spinor}
\end{equation}
where $\Omega$ is the spinor frequency, $M$ is a half-integer parameter (the azimuthal number). 
For our purposes, it is convenient to represent the components of the spinor appearing in~\eqref{spinor} 
in the following form:
$$
	\psi_1=\frac{1}{2}\left[X+Y+\imath\left(V+W\right)\right],\quad
	\psi_2=\frac{1}{2}\left[X-Y+\imath\left(V-W\right)\right],
$$
where the functions $X,Y,V$, and $W$ depend only on the spherical coordinates $r$ and $\theta$,
and the line element is
$$
	ds^2=dt^2-dr^2-r^2\left(d\theta^2+\sin^2\theta d\varphi^2\right) .
$$

The {\it Ansatz} for the Maxwell field is taken to be 
\begin{equation}
	A_\mu = \{\phi(r,\theta), 0, 0, \sigma(r,\theta)\} ,
\label{EM_ans}
\end{equation}
i.e., it contains an electric and a magnetic potentials. This {\it Ansatz} implies the presence of the following nonzero components of the electric and magnetic fields:
\begin{equation}
	E_r = - \frac{\partial \phi}{\partial r}, \quad 
	E_\theta = - \frac{\partial \phi}{\partial \theta}, \quad
	H_r = - \frac{\csc\theta}{r^2} \frac{\partial \sigma}{\partial \theta}, \quad 
	H_\theta = \csc\theta \frac{\partial \sigma}{\partial r} .
\label{EM_components}
\end{equation}

\section{Equations and solutions}

Substituting the {\it Ans\"{a}tze} \eqref{spinor} and \eqref{EM_ans} in the field equations~\eqref{feqs-20} and \eqref{feqs-22}, one can obtain the following set of six partial differential equations:
\begin{align}
	\tilde X_{,x} + \frac{\tilde X}{x} - \frac{\tilde W_{,\theta}}{x} 
  - \frac{\cot\frac{\theta}{2}}{2x} \tilde W 
  + \tilde Q
  \left(
  	- \frac{\csc\theta }{x}\tilde W \tilde \sigma 
   + \tilde V \tilde \phi
   \right) - \left(1 + \tilde \Omega\right)\tilde V 
   + U_2 \tilde V & = 0 ,
\label{Dirac_eq_X}\\
   \tilde Y_{,x} + \frac{\tilde Y}{x} - \frac{\tilde V_{,\theta}}{x} 
   + \frac{\tan\frac{\theta}{2}}{2x} \tilde V 
   + \tilde Q 
   \left(
   		\frac{\csc\theta }{x}\tilde V \tilde \sigma - \tilde W \tilde \phi
   	\right) - \left( 1 - \tilde \Omega\right)\tilde W 
   	+ U_2 \tilde W&=0 ,
\label{Dirac_eq_Y}\\
    \tilde V_{,x} + \frac{\tilde V}{x} + \frac{\tilde Y_{,\theta}}{x} 
    + \frac{\cot\frac{\theta}{2}}{2x}\tilde Y 
    + \tilde Q
    \left(
    	\frac{\csc\theta }{x}\tilde Y \tilde \sigma-\tilde X \tilde \phi
    \right) - \left(1 - \tilde \Omega\right) \tilde X 
    + U_2 \tilde X & = 0 ,
\label{Dirac_eq_V}\\
   \tilde W_{,x} + \frac{\tilde W}{x} + \frac{\tilde X_{,\theta}}{x} 
   - \frac{\tan\frac{\theta}{2}}{2x} \tilde X 
   + \tilde Q
   \left(
   	- \frac{\csc\theta }{x}\tilde X \tilde \sigma + \tilde Y \tilde \phi
   \right)-\left(1+\tilde \Omega\right)\tilde Y
   + U_2 \tilde Y & = 0 ,
\label{Dirac_eq_W}\\
   \tilde \phi_{,xx} + \frac{2}{x}\tilde \phi_{,x} + 
   \frac{1}{x^2}\tilde \phi_{,\theta\theta} 
   + \frac{\cot\theta}{x^2}\tilde \phi_{,\theta} 
   + \tilde Q\, U_1 & = 0 ,
\label{Maxw_eq_phi}\\
  \tilde \sigma_{,xx} + \frac{1}{x^2}\tilde \sigma_{,\theta\theta} 
  - \frac{\cot\theta}{x^2}\tilde \sigma_{,\theta} 
  + 2\, \tilde Q\, x \sin\theta\, U_3 & = 0 ,
\label{Maxw_eq_sigma}
\end{align}
where 
$$
	U_1 = \tilde X^2 + \tilde Y^2 + \tilde V^2 + \tilde W^2, \quad 
	U_2 = \tilde X^2 - \tilde Y^2 - \tilde V^2 + \tilde W^2, \quad 
	U_3 = \tilde X \tilde Y + \tilde V \tilde W
$$ 
and the azimuthal number is taken to be $M=1/2$ throughout the paper. These equations are written in terms of the following dimensionless variables:
$x=m r$, $\tilde\Omega=\Omega/m$, 
$\tilde Q=Q/\left(m\sqrt{\lambda}\right)$, 
$\tilde X, \tilde Y, \tilde V, \tilde W=\sqrt{\lambda/m}\,X,Y,V,W$,
$\tilde{\phi}=\sqrt{\lambda}\,\phi$, 
$\tilde{\sigma}=m\sqrt{\lambda}\,\sigma$. 
The lower indices denote differentiation with respect to the corresponding coordinate. Notice that these equations do not explicitly contain the nonlinearity parameter $\lambda$ 
and are invariant with respect to multiplying the spinor functions by $-1$. That is, the system contains only two free parameters, $\tilde \Omega$ and $\tilde Q$, 
whose values will be varied to obtain solutions describing configurations with different physical characteristics. 

\subsection{Physical quantities}

Let us now write down expressions for some physically interesting parameters of the systems under consideration. The total dimensionless mass of the system can be found in the form
\begin{equation}
	\tilde{M}_{\text{tot}}\equiv m\lambda M_{\text{tot}} = 
	2 \pi \int_{0}^{\infty} dx \int_{0}^{\pi} d \theta\,\tilde{T}_t^t x^2 \sin\theta  ,
\label{mass_tot}
\end{equation}
where the dimensionless $(^t_t)$-component of the energy-momentum tensor~\eqref{EM} is
\begin{equation}
	\tilde{T}_t^t = \frac{1}{2}\left[
		\tilde \phi_{,x}^2 + \frac{\csc^2\theta }{x^2}\tilde \sigma_{,x}^2 
		+ \frac{1}{x^2}\tilde \phi_{,\theta}^2 
		+ \frac{\csc^2\theta }{x^4}\tilde \sigma_{,\theta}^2 
		+ 2 \left(\tilde \Omega - \tilde Q \tilde \phi\right) U_1 
		+ U_2^2
	\right] .
\label{T_tt}
\end{equation}

The total dimensionless angular momentum
\begin{equation}
\tilde{J}_{\text{tot}}\equiv m^2\lambda J_{\text{tot}}=-2\pi \int_{0}^{\infty} dx \int_{0}^{\pi} d\theta\, \tilde{T}_\varphi^t x^2 \sin\theta  ,
\label{ang_mom_tot}
\end{equation}
where the dimensionless $(^t_\varphi)$-component of the energy-momentum tensor~\eqref{EM} is
\begin{equation}
\begin{split}
\tilde{T}_\varphi^t=&\tilde \phi_{,x}\tilde \sigma_{,x}+\frac{1}{x^2}\tilde \phi_{,\theta}\tilde \sigma_{,\theta}-\frac{1}{4}\left(1+2\, \tilde Q\, \tilde \sigma\right) U_1+x\sin\theta \left(\tilde \Omega-\tilde Q \tilde \phi\right) U_3 \\
&+\frac{1}{2}\sin\theta\left(\tilde W \tilde X-\tilde V \tilde Y\right)+\frac{1}{4}\cos\theta \left(\tilde X^2-\tilde Y^2+\tilde V^2-\tilde W^2\right) .
\label{T_tphi}
\end{split}
\end{equation}
The occurrence of a nonzero angular momentum is due to the presence in the system of (i) a single fermion possessing an intrinsic angular momentum;
and (ii) the crossed electric and magnetic fields. For this reason, in analogy to quantum particles possessing the quantum-mechanical spin, such configurations can be treated as spinning ones.

The total dimensionless Noether charge
\begin{equation}
\tilde{Q}_{\text{tot}}\equiv m^2\lambda\, Q_{\text{tot}}=2\pi \int_{0}^{\infty} dx \int_{0}^{\pi} d\theta\, \tilde{j}^t x^2 \sin\theta  ,
\label{Noether}
\end{equation}
where the dimensionless temporal component of the current density $\tilde{j}^t= U_1$.
Note here that the normalization condition $Q_{\text{tot}}=1$ corresponds to one-particle solutions; 
in this case the coupling constant $Q$ will correspond to an electric charge of the system, and below we will be interested mostly in such configurations.

Magnetic moment of the system under investigation can be calculated in a standard way by considering the electric current flowing perpendicular to a meridional
plane (see, e.g., the textbook~\cite{Blokh}). As a result, one can obtain the following expression for the 
dimensionless magnetic dipole moment:
\begin{equation}
\tilde{\mu}_m \equiv m^2\sqrt{\lambda}\,\mu_m=-2\pi \tilde Q \int_{0}^{\infty} dx \int_{0}^{\pi} d\theta\, U_3 x^3 \sin^2\theta .
\label{mag_mom}
\end{equation}

Finally, the dimensionless gyromagnetic ratio $g$, expressed in units of $q_e/\left(2 M_{\text{tot}}\right)$ [where $q_e$ is the electric charge, see Eq.~\eqref{asympt_behav} below], 
is defined from the relation
\begin{equation}
\mu_m=g\frac{ q_e}{2}\frac{ J_{\text{tot}}}{M_{\text{tot}}} \quad \Rightarrow \quad g=2\frac{\tilde{\mu}_m \tilde{M}_{\text{tot}}}{\tilde{q}_e \tilde{J}_{\text{tot}}},
\label{gyro}
\end{equation}
where $\tilde{q}_e\equiv m\sqrt{\lambda} \,q_e$ is the dimensionless electric charge.

\subsection{Boundary conditions and a numerical approach}

We will seek globally regular finite-energy nodeless solutions of the set of six partial differential equations \eqref{Dirac_eq_X}-\eqref{Maxw_eq_sigma}. 
To do this, it is necessary to impose appropriate boundary conditions
for the spinor and Maxwell fields. The behavior of solutions of Eqs.~\eqref{Dirac_eq_X}-\eqref{Maxw_eq_sigma} 
in the vicinity of the boundaries of the domain of integration implies the following boundary conditions:
\begin{equation}
\begin{split}
&\left. \frac{\partial \tilde X}{\partial x}\right|_{x = 0} =
	\left. \frac{\partial \tilde W}{\partial x}\right|_{x = 0} =
    \left. \frac{\partial \tilde \phi}{\partial x}\right|_{x = 0}=  0,  \left. \tilde Y \right|_{x = 0} =\left. \tilde V \right|_{x = 0}=\left. \tilde \sigma \right|_{x = 0}= 0;\\
&\left. \tilde X \right|_{x = \infty} = 
	\left. \tilde Y \right|_{x = \infty} = 
    \left. \tilde V \right|_{x = \infty} = 
    \left. \tilde W \right|_{x = \infty} = 
	\left. \tilde \phi \right|_{x = \infty} =
    \left. \tilde \sigma \right|_{x = \infty} =   0 ; \\
&\left. \frac{\partial \tilde X}{\partial \theta}\right|_{\theta = 0} =
    \left. \frac{\partial \tilde V}{\partial \theta}\right|_{\theta = 0} =
	\left. \frac{\partial \tilde \phi}{\partial \theta}\right|_{\theta = 0} =  0 ,  \left. \tilde Y \right|_{\theta = 0} =\left. \tilde W \right|_{\theta = 0}=\left. \tilde \sigma \right|_{\theta = 0}= 0 ;\\
&\left. \frac{\partial \tilde Y}{\partial \theta}\right|_{\theta = \pi} =
	\left. \frac{\partial \tilde W}{\partial \theta}\right|_{\theta = \pi} =
	\left. \frac{\partial \tilde \phi}{\partial \theta}\right|_{\theta = \pi} =  0 ,  \left. \tilde X \right|_{\theta = \pi} =\left. \tilde V \right|_{\theta = \pi}=\left. \tilde \sigma \right|_{\theta = \pi}= 0.
\label{BCs}
\end{split}
\nonumber
\end{equation}
For numerical computations, it is convenient to introduce the compactified radial coordinate
\begin{equation}
	\bar x = \frac{x}{1+x} 
\label{comp_coord}
\end{equation}
in order to map the infinite interval $[0,\infty)$ to the finite region $[0,1]$. The results of numerical computations for axisymmetric systems presented below 
have been obtained using the Intel MKL PARDISO sparse direct solver and the CESDSOL library, and also verified for some particular cases using the package FIDISOL~\cite{fidisol}. 
These packages provide  an iterative procedure for obtaining an exact solution starting from some approximate solution (an initial guess). As the initial guess, it is possible to use the solutions
describing configurations in the absence of electric and magnetic fields~\cite{Soler:1970xp}. The equations~\eqref{Dirac_eq_X}-\eqref{Maxw_eq_sigma} have been solved on a grid of $200\times 100$ 
points which covers  the integration region  $0\leq \bar x \leq 1$ [given by the compactified radial coordinate~\eqref{comp_coord}] and $0\leq \theta \leq \pi$.

\subsection{Numerical solutions}

\begin{figure}[t!]
\begin{center}
\includegraphics[width=.6\linewidth]{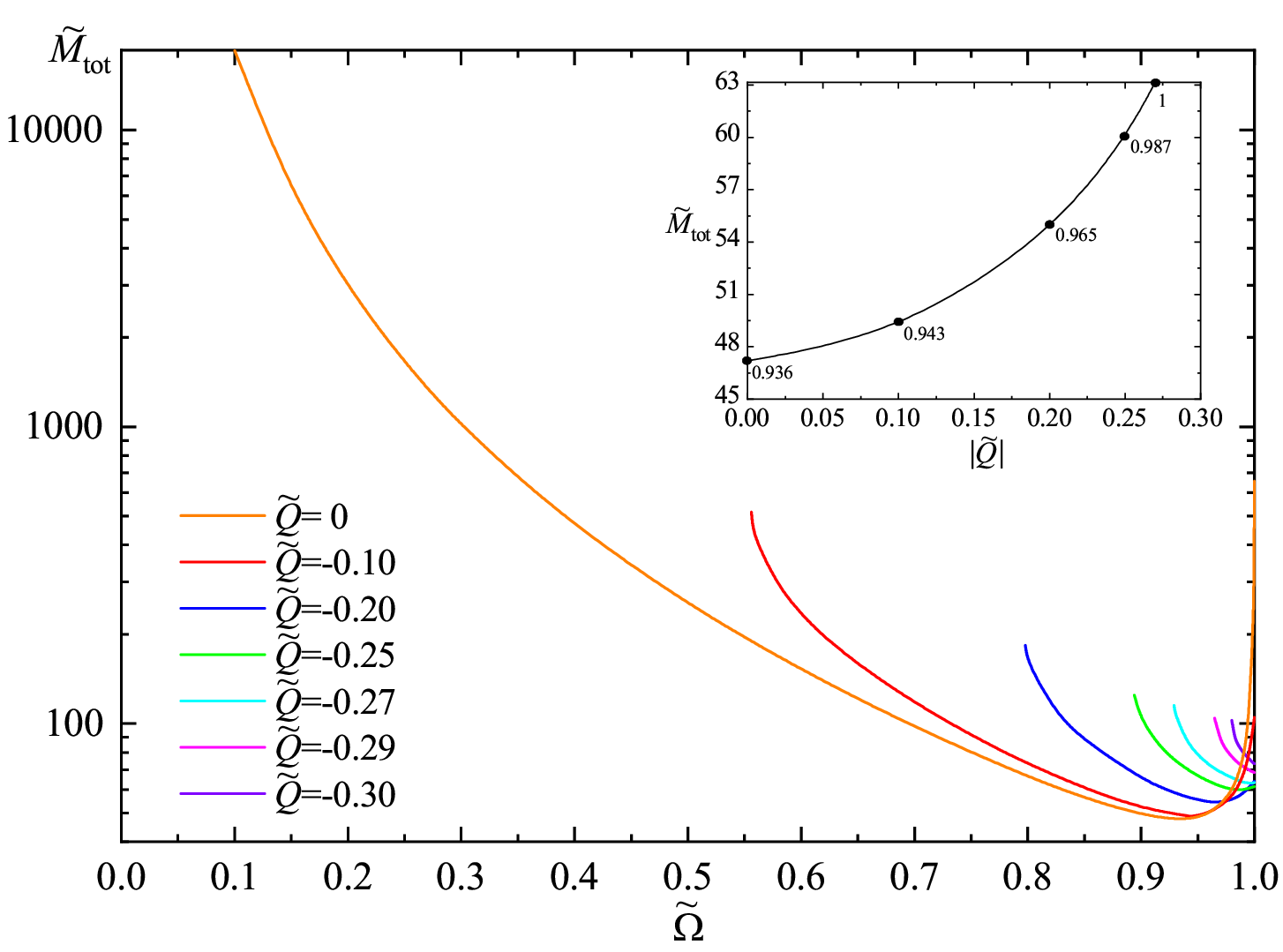}
\end{center}
\vspace{-0.5cm}
\caption{
	The total mass of the system $\tilde M_{\text{tot}}$ as a function of the spinor frequency  $\tilde{\Omega}$ 
	for different values of the coupling constant $\tilde Q$. The inset shows the dependence of the total mass on  $\tilde Q$ 	for different values of $\tilde\Omega$ (shown by the numbers near the points) corresponding to the location of the mass gap. 
}
\label{fig_Mass_Omega}
\end{figure}

In contrast to the case of a linear spinor field, in the nonlinear case, there is a family of solutions, depending continuously on two parameters~-- the frequency  $\tilde{\Omega}$ 
and the coupling constant $\tilde Q$, whose values completely determine all physical characteristics of the configurations under consideration.
To illustrate this, Fig.~\ref{fig_Mass_Omega} shows the spectrum of the total mass \eqref{mass_tot} of the systems under investigation as a function of $\tilde{\Omega}$ 
for some fixed values of $\tilde Q$. It is seen from these graphs that the behavior of the dependence  $\tilde M_{\text{tot}}(\tilde{\Omega})$ is largely determined  
by the value of the coupling constant~$\tilde Q$. Namely, the numerical calculations indicate that: 
\begin{itemize}
\item[(i)] When $\tilde Q=0$, the total mass diverges as $\tilde{\Omega}\to 1$. In turn, for small $\tilde{\Omega}$, a rapid increase of $\tilde M_{\text{tot}}$ also occurs, 
and one may expect that in the limit $\tilde{\Omega}\to 0$ the total mass will tend to infinity as well. But in this limit a calculation of the mass using the integral~\eqref{mass_tot} 
is a difficult technical problem, and we cannot verify this assumption by direct calculation.

\item[(ii)] When $\tilde Q\neq 0$, the total mass in the limit $\tilde{\Omega}\to 1$ is already finite. In turn, there is some nonzero value  $\tilde{\Omega}<1$ 
for which one can still perform numerical calculations. In doing so, one can observe that the mass demonstrates a rapid increase ($|\partial \tilde M_{\text{tot}}/\partial \tilde{\Omega}| \gg 1$); 
this can be regarded as an indication that there is some critical value   $\tilde{\Omega}_{\text{crit}}$ for which the mass will eventually tend to infinity.

\item[(iii)] A distinctive feature of the configurations from (i) and (ii) is the presence of a minimum of the mass for all values of $\tilde Q$ lying in the interval $-0.27\lesssim \tilde Q \leq 0$. 
This minimum corresponds to the presence in the system of a mass gap where $\partial \tilde M_{\text{tot}}/\partial \tilde{\Omega} =0$. In turn, for $ \tilde Q \lesssim -0.27$, 
such a mass gap is already absent: the total mass demonstrates a gradual decrease as $\tilde{\Omega}$ increases, and eventually $\tilde M_{\text{tot}}$ reaches some minimum value as $\tilde{\Omega}\to 1$.

\item[(iv)] The dependence of the total mass on the coupling constant $\tilde Q$ for different values of $\tilde\Omega$ corresponding to the location 
of the mass gap is shown in the inset of Fig.~\ref{fig_Mass_Omega}. It is seen from this inset that for the system with $\tilde Q=0$ (and correspondingly without 
the magnetic field), $\tilde{\Omega}\approx 0.936$ (cf. Ref.~\cite{Ranada:1973hna}). On the other hand, there exists a maximum possible value of the coupling 
constant $|\tilde{Q}|\approx 0.27$ for which the frequency $\tilde{\Omega}\to 1$. For larger (modulus) values of $\tilde{Q}$ the curve $\tilde M_{\text{tot}}(\tilde\Omega)$ 
has no minimum already, that is, the derivative $\partial \tilde M_{\text{tot}}/\partial \tilde\Omega$ is nowhere equal to zero, and correspondingly the mass gap is absent.

\item[(v)] There is some critical value of the coupling constant $\tilde{Q}_{\text{crit}}$ approaching that the interval $\tilde{\Omega}_{\text{crit}}\leq\tilde{\Omega}\leq 1$ (where the solutions do exist) becomes narrower, 
and eventually $\tilde{\Omega}_{\text{crit}}\to 1$ and the complete set of solutions with different $\tilde{\Omega}$ degenerates to the only solution with $\tilde{\Omega}=1$.  
Numerical calculations show that $\tilde{Q}_{\text{crit}}\approx -0.3119$, and in this limit the total mass $\tilde M_{\text{tot}}\approx 97.13$.

\item[(vi)] For all these spinning systems, a straightforward computation shows that the total angular momentum $J_{\text{tot}}$ from Eq.~\eqref{ang_mom_tot} 
and the total Noether charge $Q_{\text{tot}}$ from Eq.~\eqref{Noether} are related by $ J_{\text{tot}}=\frac{1}{2}  Q_{\text{tot}}$, although the angular momentum density and the Noether charge density are not proportional.

\item[(vii)] From the form of Eqs.~\eqref{Dirac_eq_X}-\eqref{Maxw_eq_sigma}, it is evident that the solutions under consideration are invariant with respect 
to a change in the sign $\tilde Q\to -\tilde Q, \tilde\phi\to -\tilde\phi, \tilde\sigma \to -\tilde\sigma$, that is, they may describe systems with the coupling constant opposite in sign and the same physical characteristics.
\end{itemize}

Note here that, in order to apply the results obtained above for a description of one particle, it is necessary to normalize the solutions so that the 
Noether charge $Q_{\text{tot}}$ from Eq.~\eqref{Noether} would be equal to~1. In this case the coupling constant $Q$ will correspond to an electric charge, i.e., $Q=q_e$. 
This one-particle condition can be fulfilled by a suitable choice of the free system parameters $\lambda$ and $m$. However, in doing so, one should bear in mind that in 
this case there will be its own particular set of the parameters $\lambda$ and $m$ for every point in the $\left[\tilde M_{\text{tot}}-\tilde{\Omega}\right]$-plane,
i.e., different points of the plane will correspond to different models. 

We conclude this subsection with the expression for the effective radial pressure $p_r\equiv -T_r^r$. Using the energy-momentum tensor~\eqref{EM} and the Dirac equations~\eqref{Dirac_eq_X}-\eqref{Dirac_eq_W},
it can be shown that the radial pressure contains the terms
$$
p_r \sim \frac{\lambda}{2}\left(X^2- Y^2- V^2+ W^2\right)^2 -Q \phi \left(X^2+ Y^2+ V^2+ W^2\right)\ldots
$$
This implies the following physical meaning of the nonlinearity parameter: the case of  $\lambda > 0$ corresponds to the attraction and the case of $\lambda < 0$ 
to the repulsion. Correspondingly, for the configurations considered above, the attraction of the spinor field related to the choice of positive values of the nonlinearity parameter provides a counter-balance to the effective 
repulsion due to the presence of the electric charge.

\subsection{Asymptotic behavior}

For completeness of analysis of the numerical solutions obtained above, let us write down analytical expressions for asymptotic solutions. 
The Maxwell equations~\eqref{Maxw_eq_phi} and \eqref{Maxw_eq_sigma} have the following asymptotic
($x\to \infty$) behavior of the electric and magnetic fields:
\begin{equation}
	\tilde \phi \approx \frac{1}{4\pi} \frac{\tilde{q}_e}{x} , \quad 
	\tilde \sigma \approx - \frac{1}{4\pi}\frac{\tilde{\mu}_m}{x} \sin^2 \theta .
\label{asympt_behav}
\end{equation}
 Using these expressions,  the numerical values of the electric charge $\tilde{q}_e$ and of the magnetic moment $\tilde{\mu}_m$ can be found in the form
\begin{equation}
	\tilde{q}_e = - 4 \pi \lim_{x\to\infty} x^2 \frac{\partial\tilde \phi}{\partial x} 
	= - 4 \pi \lim_{\bar{x}\to 1} \bar{x}^2 \frac{\partial\tilde \phi}{\partial\bar{x}}, \quad
	\tilde{\mu}_m = 4 \pi \lim_{x\to\infty} \frac{x^2}{\sin^2\theta} 
	\frac{\partial\tilde \sigma}{\partial x} 
	= 4 \pi \lim_{\bar{x}\to 1} \frac{\bar{x}^2}{\sin^2\theta} \frac{\partial\tilde \sigma}{\partial\bar{x}} .
\label{asympt_charge}
\end{equation}
Note that the value of $\tilde{\mu}_m$ calculated using the above formula coincides with that of obtained using Eq.~\eqref{mag_mom}. 

In turn, the asymptotic behavior of the spinor fields follows from the Dirac equations ~\eqref{Dirac_eq_X}-\eqref{Dirac_eq_W},
$$
	\tilde X\approx -2 \cos\frac{\theta}{2}\, g(x), \quad 
	\tilde Y\approx 2 \sin\frac{\theta}{2}\, f(x), \quad 
	\tilde V\approx 2 \cos\frac{\theta}{2}\, f(x), \quad 
	\tilde W\approx -2 \sin\frac{\theta}{2}\, g(x) .
$$
The form of the functions $f(x)$ and $g(x)$ appearing here depends on the value of $\tilde{\Omega}$. Namely, for $0<\tilde{\Omega}<1$, we have
\begin{equation}
	f(x)\approx f_\infty \frac{e^{-\sqrt{1-\tilde\Omega^2}\,x}}{x} x^{-\frac{1}{\sqrt{1-\tilde\Omega^2}}\frac{\tilde{Q}\tilde{q}_e}{4\pi}}, \quad
g(x)\approx f_\infty \sqrt{\frac{1+\tilde\Omega}{1-\tilde\Omega}}\,\frac{e^{-\sqrt{1-\tilde\Omega^2}\,x}}{x} x^{-\frac{1}{\sqrt{1-\tilde\Omega^2}}\frac{\tilde{Q}\tilde{q}_e}{4\pi}},
	\label{asympt}
\end{equation}
where $f_\infty$ is an integration constant. In the case of $\tilde{\Omega}=1$, we have 
$$
	f(x) \approx f_\infty \frac{e^{-\sqrt{\frac{2}{\pi}\tilde{Q}\tilde{q}_e\,x}}}{x^{5/4}}, \quad
	g(x) \approx f_\infty \sqrt{\frac{8\pi}{\tilde{Q}\tilde{q}_e}}\,
	\frac{e^{-\sqrt{\frac{2}{\pi}\tilde{Q}\tilde{q}_e\,x}}}{x^{3/4}} .
$$
Notice here that regular solutions with $\tilde{\Omega}=1$ are only possible in the presence of the charge.

\begin{figure}[t!]
\begin{center}
\includegraphics[width=1.\linewidth]{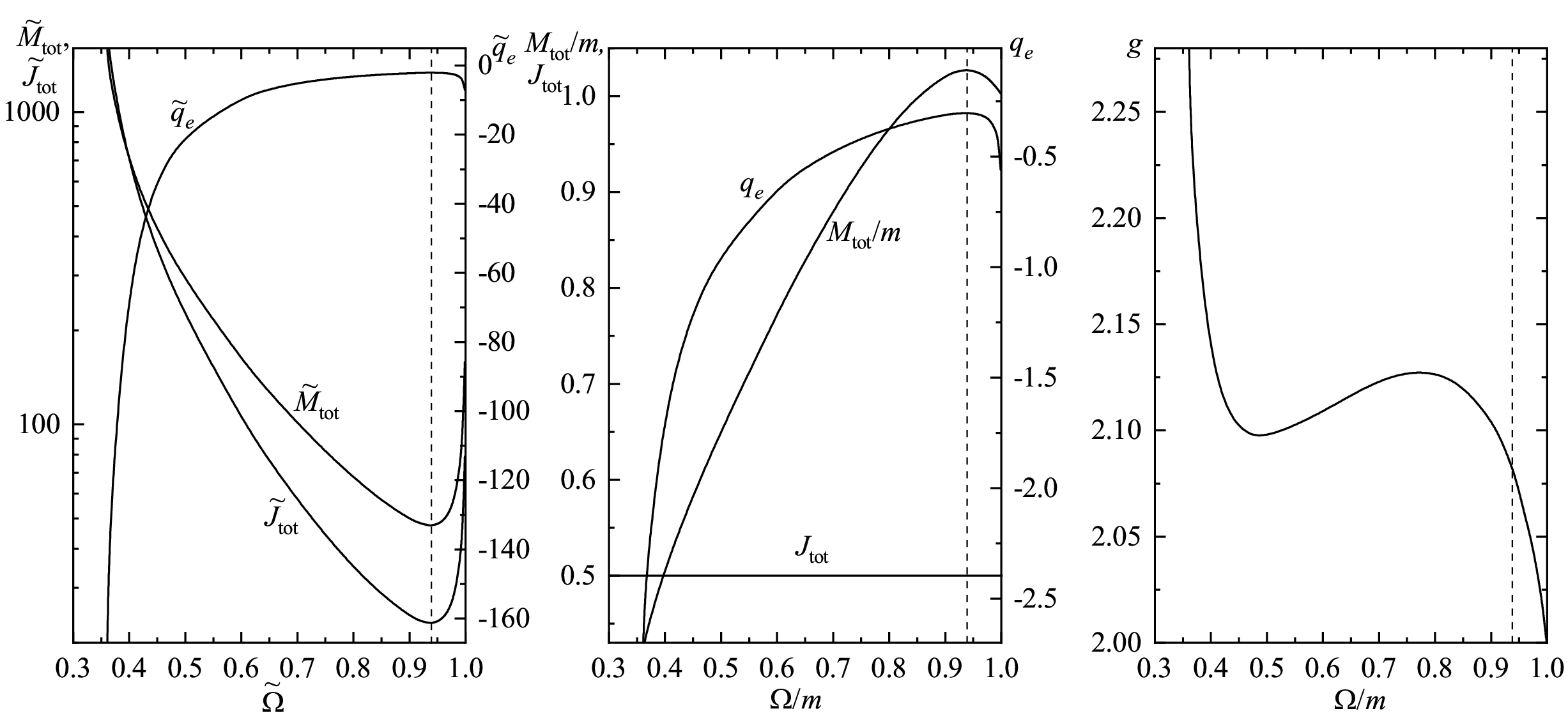}
\end{center}
\vspace{-0.5cm}
\caption{
The dependence of the physical quantities on the spinor frequency $\Omega$ for a fixed $\tilde{Q}=-0.04452$. Left panel: the graphs for the total mass 
$\tilde M_{\text{tot}}$ from Eq.~\eqref{mass_tot}, for the angular momentum $\tilde J_{\text{tot}}$ from Eq.~\eqref{ang_mom_tot},
and for the charge $\tilde{q}_e$ from Eq.~\eqref{asympt_charge}. Middle panel: the same quantities, but in the dimensional form and with $Q_{\text{tot}}=1$ 
(normalized values). Right panel: the gyromagnetic ratio~\eqref{gyro}. The vertical dashed lines correspond to the minimum of the mass (the mass gap) 
located at the point $\tilde{\Omega}\approx 0.937$.
}
\label{fig_Mass_charge}
\end{figure}

\subsection{Particular example:  an ``electron''}
\label{electron}

\begin{figure}[t!]
\begin{center}
\includegraphics[width=.8\linewidth]{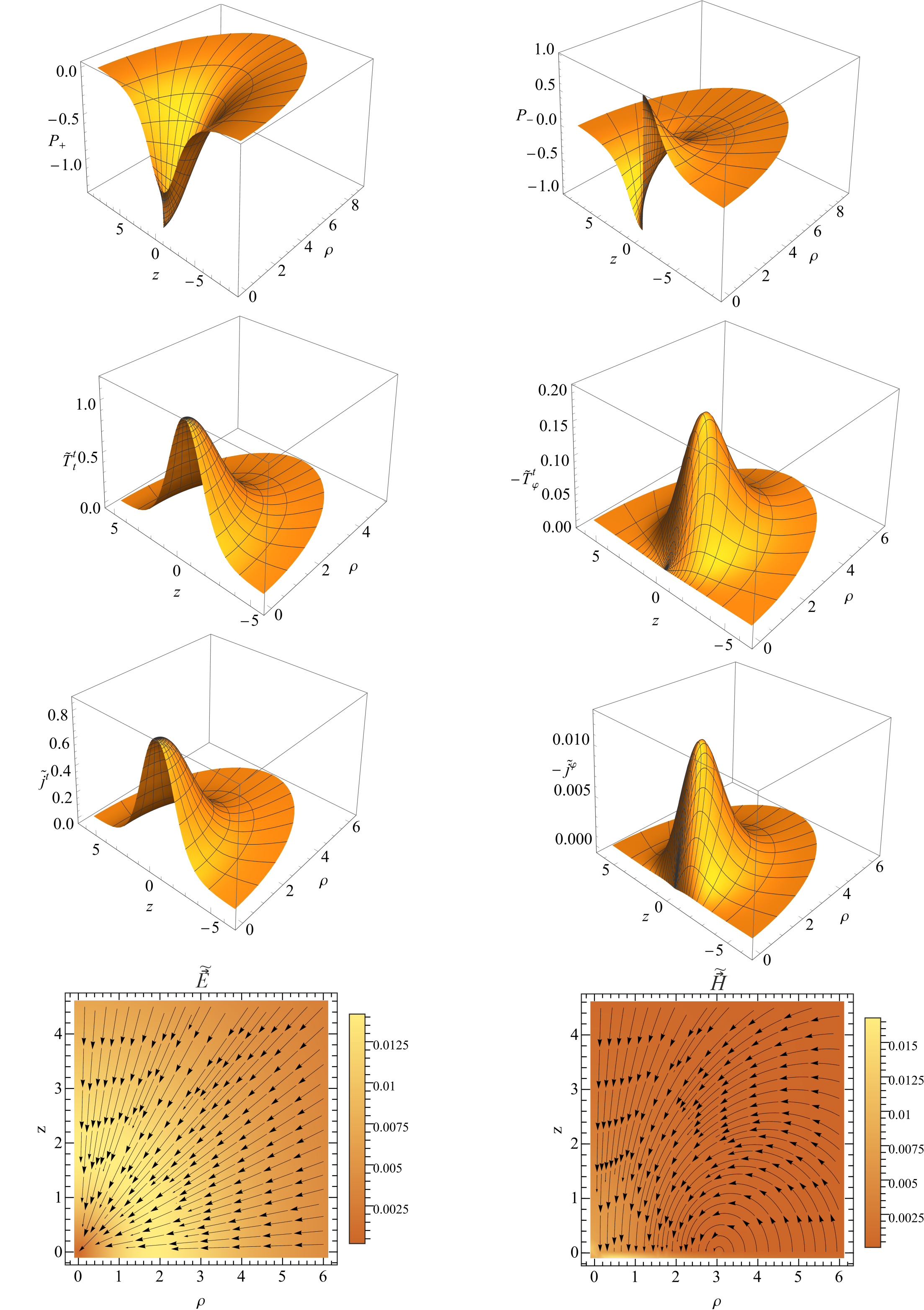}
\end{center}
\vspace{-0.5cm}
\caption{
	The dimensionless spinor functions  $P_+$ and $P_-$ from~\eqref{symm_P} and \eqref{anti_symm_P},
energy density~\eqref{T_tt}, angular momentum density~\eqref{T_tphi}, charge density $\tilde{j}^t= U_1$, physical component of the current density 
$\tilde{j}^\varphi=-2\, \tilde{Q}\, U_3$, electric, $\tilde{\vec{E}}\equiv \sqrt{\lambda}/m\vec{E}$, and magnetic, $\tilde{\vec{H}}\equiv \sqrt{\lambda}/m\vec{H}$, 
field strengths for the system with $\tilde{Q}=-0.04452$ located at the mass gap (cf. Fig.~\ref{fig_Mass_charge}). The plots are made in a meridional plane $\varphi=\text{const.}$ 
spanned by the coordinates $\rho=x \sin\theta$ and $z=x \cos\theta$. Since the system is $\mathbb{Z}_2$-symmetric with respect to the equatorial plane $z=0$, the electromagnetic field strength distributions
are shown only for $z > 0$.
}
\label{fig_T_j}
\end{figure}

The freedom in the choice of values of the parameters $\lambda$ and $m$ enables us to model various objects. In doing so, a choice of a specific value of $\tilde \Omega$ 
can be made on the basis of a physically reasonable assumption that an energetically stable system must possess a minimum energy (or, equivalently, a minimum mass 
$\tilde M_{\text{tot}}$). Consistent with this, consider, for example, the case where the coupling constant $\tilde Q$ is understood to be so chosen that at a minimum of the curve
$\tilde M_{\text{tot}}(\tilde\Omega)$ the electric charge of the system $q_e$ would be equal to the charge of an electron. The corresponding dependencies  
$\tilde M_{\text{tot}}(\tilde\Omega)$ and $\tilde{q}_e(\tilde\Omega)$ are shown in the left panel of Fig.~\ref{fig_Mass_charge}. 
It is seen that the minimum of the mass curve (the maximum of the charge curve) is located at $\tilde{\Omega} \approx 0.937$, and the total mass and Noether charge are
$$
	M_{\text{tot}}=\frac{47.512}{m\lambda}, \quad Q_{\text{tot}}=\frac{46.255}{m^2\lambda}.
$$ 
This yields the dimensional mass $M_{\text{tot}}=1.027 \,m \,Q_{\text{tot}}$ (see the middle panel of Fig.~\ref{fig_Mass_charge}). 
For a normalized solution, $Q_{\text{tot}}=1$; correspondingly, there is a mass renormalization of $2.7\%$. Then, in order to make the total mass of the system $M_{\text{tot}}$ 
equal to the electron mass $m_e$, it is necessary to take $m=m_e/1.027$. This in turn leads to the corresponding renormalization 
of the magnetic moment and change in the value of the gyromagnetic ratio $g$, whose graph is shown in the right panel of Fig.~\ref{fig_Mass_charge}. 
For the case under consideration,  $g\approx 2.083$. Thus we have a configuration with the mass $M_{\text{tot}} = 0.511\, \text{MeV}$ and charge $q_e = -0.3028$ equal to the mass and charge of an electron, but with $g > 2$. 

Also, for a normalized solution, i.e., when $Q_{\text{tot}}=1$, the quantum-mechanical angular momentum,
which is determined  using the operator of the total angular momentum,
\begin{equation}
	\hat{M}_z = \hat{L}_z + \hat S_z , 
\label{oper_TAM}
\end{equation}
(here $\hat{L}_z = - \imath \partial_\varphi$ is the operator which projects the orbital angular momentum on the $z$-axis and $ \hat S_z$ 
is the operator which projects the spin on the $z$-axis), defines the value of the total angular momentum as an eigenvalue $M_z$,
$$
	\hat{M}_z \psi = M_z \psi . 
$$
For the spinor~\eqref{spinor}, this eigenvalue is $M_z = 1/2$, and it coincides with that calculated using the integral formula~\eqref{ang_mom_tot} 
(cf. the value of $J_{\text{tot}}$ from the middle panel of Fig.~\ref{fig_Mass_charge}). It is worth noting that for the solutions that are not normalized to unity this coincidence no longer occurs.  

The characteristic size of such a charged configuration supported by the spinor field can be estimated from the asymptotic behavior of the field~\eqref{asympt} as
$$
r_{\text{ch}}\sim \frac{1}{\sqrt{1-\tilde\Omega^2}\, m_e}.
$$
For $\tilde{\Omega}\approx 0.937$, this yields $r_{\text{ch}}\sim 10^{-10}\,\text{cm}$, a value that is comparable in order of magnitude to the electron Compton wavelength. 

Note that the spinor functions $\tilde{X},\tilde{Y},\tilde{V},$ and $\tilde{W}$ appearing in the Dirac equations \eqref{Dirac_eq_X}-\eqref{Dirac_eq_W} 
are neither even nor odd functions with respect to the equatorial plane $\theta=\pi/2$. Nevertheless, the system possesses a $\mathbb{Z}_2$ 
symmetry with respect to this plane; this can be shown by considering the corresponding combinations of the spinor functions (see Appendix~\ref{append1}). 
To demonstrate this fact in a pictorial way, the upper row of Fig.~\ref{fig_T_j} shows the graphs of the functions  $P_+$ and $P_-$ from Eqs.~\eqref{symm_P} and
\eqref{anti_symm_P}, respectively. Also, this figure shows the corresponding $\mathbb{Z}_2$-symmetric distributions of the components of the energy-momentum tensor~\eqref{EM} and current density, 
as well as the electric and magnetic field strengths defined by the expressions~\eqref{EM_components}.
The structure of the magnetic field strength corresponds to an axially symmetric dipole field sourced by
the current associated with the spinor field given on the right-hand side of Eq.~\eqref{feqs-22}.
The radial distribution of the current and the magnitude of the magnetic field are determined by the value of the coupling constant $Q$. 
In turn, the structure of the electric field strength corresponds to a negative charge with force lines directed toward the center of the configuration.

In conclusion, note that by choosing $m=m_{\mu}/1.027$, where $m_{\mu}$ is the muon mass, we get characteristics typical for a muon/antimuon.

\section{Conclusions}

The main purpose of the present paper is to study self-consistently the influence that an electromagnetic field has on a system supported by a nonlinear spinor field. 
To this end, we generalized the configurations considered in Refs.~\cite{Soler:1970xp,Ranada:1973hna} by including nonperturbatively electric and magnetic (dipole) 
Maxwell fields  to take account of their backreaction on the physical characteristics of the system. In such a generalized case, the presence in the system of the dipole magnetic field requires a consideration 
of an axisymmetric problem.

In the absence of electromagnetic fields, an important distinctive feature of the systems supported by nonlinear spinor fields is the presence of a mass gap. 
For such systems, all solutions are parameterized by one parameter~-- the spinor frequency $\tilde{\Omega}$, and regular stationary solutions describing 
configurations with finite values of various physical parameters (for instance, of a total mass) do exist only for the values of $\tilde{\Omega}$ lying in the range $0<\tilde{\Omega}<1$,  
whereas for $\tilde{\Omega}\to 0$ and $\tilde{\Omega}\to 1$ the total mass diverges. The inclusion of the Maxwell fields results in the appearance of one more 
free parameter~-- the coupling constant $\tilde Q$. For such a two-parametric system, we have considered all permissible values of the parameters $\tilde{\Omega}$ and $\tilde Q$
for which regular spinning solutions do exist. The studies of the present work indicate that there are 
the following qualitative changes in the characteristics of the configurations as compared with the electrically neutral ($\tilde Q=0$) case:
\begin{itemize}
\item Apart from the mass and angular momentum gaps, the system also contains the charge and magnetic moment gaps located at the same values of $\tilde{\Omega}$ as the mass gap 
(see Fig.~\ref{fig_Mass_charge} and cf. Ref.~\cite{Dzhunushaliev:2022pkz}). 
\item For a nonzero coupling constant $\tilde Q$, there is some critical value of the spinor frequency $\tilde{\Omega}_{\text{crit}}>0$ that restricts 
the range of permissible values of $\tilde{\Omega}$ from the left. As in the case without Maxwell fields, at this boundary, the total masses of the system diverge 
for all permissible values of $\tilde Q$. As $\tilde Q$ increases (modulus), the value of $\tilde{\Omega}_{\text{crit}}$ increases as well, and there is a finite critical 
value  $|\tilde{Q}_{\text{crit}}|\approx 0.3119$ for which  $\tilde{\Omega}_{\text{crit}}\to 1$, i.e., the set of solutions with different $\tilde{\Omega}$ 
degenerates to the only solution with $\tilde{\Omega}=1$. 
\item For $0<|\tilde{Q}|<|\tilde{Q}_{\text{crit}}|$ and as $\tilde{\Omega}\to 1$, the total mass of the system, in contrast to the case without Maxwell fields, 
remains always finite. That is, regular solutions exist in the frequency range of $\tilde{\Omega}_{\text{crit}}<\tilde{\Omega}\leq1$, and the magnitude of
$\tilde{\Omega}_{\text{crit}}$ is completely determined only by the value of the coupling constant $\tilde Q$.
\item There is a maximum possible value of $|\tilde{Q}|\approx 0.27$ above which the aforementioned gaps are already absent in the system.
\end{itemize}

As a possible application of the above results, we have considered the case where the coupling constant $\tilde Q$ is to be so chosen that the electric charge 
of the system located at the mass/charge gap would be equal to the charge of an electron (or of a positron when $-\tilde Q$ is changed into $\tilde Q$) (see Sec.~\ref{electron}). 
Then, by choosing an appropriate value of the bare mass $m$ in the Dirac equation~\eqref{feqs-20}, one can also get the total mass of the system, equal to the mass of an electron.
In turn, the total angular momentum $J_{\text{tot}}$ and the total Noether charge $Q_{\text{tot}}$ are related by $ J_{\text{tot}}=\frac{1}{2}  Q_{\text{tot}}$ 
(as it also takes place for all other spinning systems with permissible values of $\tilde Q$ and $\tilde \Omega$ considered in the present paper), and when $Q_{\text{tot}}=1$ 
(i.e., for a normalized solution), $J_{\text{tot}}$ coincides with the eigenvalue $M_z = 1/2$ of the operator of the total angular momentum~\eqref{oper_TAM}. 
Also, the gyromagnetic ratio for such configuration is $g\approx 2.083$ (cf. the electron for which $g\approx 2$) and the characteristic size is $r_{\text{ch}}\sim 10^{-10}\,\text{cm}$.

\section*{Acknowledgments}
This research was funded by the Committee of Science of the Ministry of Science and Higher Education of the Republic of Kazakhstan (Grant No.~BR21881941).

\appendix

\section{Symmetry of the field equations}
\label{append1}

When $\theta$ is replaced by $\pi - \theta$, the equations~\eqref{Dirac_eq_X}-\eqref{Maxw_eq_sigma} take the form
\begin{align}
	\tilde X_{,x} + \frac{\tilde X}{x} + \frac{\tilde W_{,\theta}}{x} 
	- \frac{\tan\frac{\theta}{2}}{2x}\tilde W 
	+ \tilde Q \left(
		- \frac{\csc\theta }{x}\tilde W \tilde \sigma + \tilde V \tilde \phi
	\right) - \left(1 + \tilde \Omega\right)\tilde V + U_2 \tilde V = & 0 ,
\label{app_Dirac_eq_X}\\
	\tilde Y_{,x} + \frac{\tilde Y}{x} + \frac{\tilde V_{,\theta}}{x} 
	+ \frac{\cot\frac{\theta}{2}}{2x}\tilde V 
	+ \tilde Q \left(
		\frac{\csc\theta }{x}\tilde V \tilde \sigma-\tilde W \tilde \phi
	\right) - \left(1 - \tilde \Omega\right)\tilde W + U_2 \tilde W = & 0 ,
\label{app_Dirac_eq_Y}\\
	\tilde V_{,x} + \frac{\tilde V}{x} - \frac{\tilde Y_{,\theta}}{x} 
	+ \frac{\tan\frac{\theta}{2}}{2x}\tilde Y 
	+ \tilde Q 
		\left( 
			\frac{\csc\theta }{x}\tilde Y \tilde \sigma - \tilde X \tilde \phi 
		\right) 
	- \left(1 - \tilde \Omega\right)\tilde X + U_2 \tilde X = & 0 ,
\label{app_Dirac_eq_V}\\
	\tilde W_{,x} + \frac{\tilde W}{x} - \frac{\tilde X_{,\theta}}{x} 
	- \frac{\cot\frac{\theta}{2}}{2x}\tilde X 
	+ \tilde Q \left(
		- \frac{\csc\theta }{x}\tilde X \tilde \sigma+\tilde Y \tilde \phi
	\right) - \left(1 + \tilde \Omega\right)\tilde Y + U_2 \tilde Y = & 0 ,
\label{app_Dirac_eq_W}\\
	\tilde \phi_{,xx} + \frac{2}{x}\tilde \phi_{,x} 
	+ \frac{1}{x^2}\tilde \phi_{,\theta\theta} 
	+ \frac{\cot\theta}{x^2}\tilde \phi_{,\theta} 
	+ \tilde Q\, U_1 = & 0 ,
\label{app_Maxw_eq_phi}\\
	\tilde \sigma_{,xx} + \frac{1}{x^2}\tilde \sigma_{,\theta\theta} 
	- \frac{\cot\theta}{x^2}\tilde \sigma_{,\theta} + 2 \, \tilde Q\, x \sin\theta\, U_3 = & 0 .
\label{app_Maxw_eq_sigma}
\end{align}
All the functions appearing here depend on $\pi - \theta$: 
$
	\tilde X(x, \pi - \theta), \tilde Y(x, \pi - \theta), 
	\tilde V(x, \pi - \theta), \tilde W(x, \pi - \theta), 
	\phi(x, \pi - \theta), \sigma(x, \pi - \theta)
$, while the combinations $U_{1}, U_{2},$ and $U_{3}$ remain unchanged. A comparison of the equations~\eqref{app_Dirac_eq_X} and \eqref{Dirac_eq_W} 
enables one to conclude that $\tilde W(x, \theta) = \tilde X(x, \pi - \theta)$. Similarly, one can show that such a symmetry is valid for all functions:
\begin{equation}
	\tilde X(x, \theta) = \tilde W(x, \pi - \theta), \quad 
	\tilde Y(x, \theta) = \tilde V(x, \pi - \theta), \quad 
	\tilde V(x, \theta) = \tilde Y(x, \pi - \theta), \quad 
	\tilde W(x, \theta) = \tilde X(x, \pi - \theta) . 
\label{symm_non_lcl} 
\end{equation}
This enables us to rewrite four local Dirac equations~\eqref{Dirac_eq_X}-\eqref{Dirac_eq_W} in the form of two nonlocal equations
\begin{equation}
\begin{split}
	\tilde X(x, \theta)_{,x} + \frac{\tilde X(x, \theta)}{x} - \frac{\tilde X(x, \pi - \theta)_{,\theta}}{x} 
	- \frac{\cot\frac{\theta}{2}}{2x}\tilde X(x, \pi - \theta)  
	+ \tilde Q \left[
	- \frac{\csc\theta }{x}\tilde X(x, \pi - \theta) \tilde \sigma 
	+ \tilde Y(x, \pi - \theta) \tilde \phi
	\right] & 
 \\
	- \left(1 + \tilde \Omega\right)\tilde Y(x, \pi - \theta) + U_2 \tilde Y(x, \pi - \theta) = & 0 ,
\\
	\tilde Y(x, \theta)_{,x} + \frac{\tilde Y(x, \theta)}{x} 
	- \frac{\tilde Y(x, \pi - \theta)_{,\theta}}{x} 
+ \frac{\tan\frac{\theta}{2}}{2x}\tilde Y(x, \pi - \theta) 
+ \tilde Q \left[
		\frac{\csc\theta }{x}\tilde Y(x, \pi - \theta) \tilde \sigma-\tilde X(x, \pi - \theta) \tilde \phi
	\right] & 
\\
	- \left(1 - \tilde \Omega\right)\tilde X(x, \pi - \theta) + U_2 \tilde X(x, \pi - \theta) = & 0 .
\end{split}
\nonumber
\end{equation}

In turn, taking into account the properties of the functions $\tilde X, \tilde Y, \tilde V$, and $\tilde W$ given in Eq.~\eqref{symm_non_lcl}, one can see that there are the following symmetric and antisymmetric functions:
\begin{align}
	P_+(x, \theta) = & \tilde X(x, \theta) + \tilde W(x, \theta) 
	= \tilde X(x, \theta) + \tilde X(x, \pi - \theta) 
	= \tilde W(x, \theta) + \tilde W(x, \pi - \theta) 
	= P_+(x, \pi - \theta) , 
\label{symm_P}\\
	P_-(x, \theta) = & \tilde X(x, \theta) - \tilde W(x, \theta) 
	= \tilde X(x, \theta) - \tilde X(x, \pi - \theta) 
	= \tilde W(x, \pi - \theta) - \tilde W(x, \theta) 
	= - P_-(x, \pi - \theta) , 
\label{anti_symm_P}\\
	Q_+(x, \theta) = & \tilde Y(x, \theta) + \tilde V(x, \theta) 
	= \tilde Y(x, \theta) + \tilde Y(x, \pi - \theta) 
	= \tilde V(x, \theta) + \tilde V(x, \pi - \theta) 
	= Q_+(x, \pi - \theta) , 
\label{symm_Q}\\
	Q_-(x, \theta) = & \tilde Y(x, \theta) - \tilde V(x, \theta) 
= \tilde Y(x, \theta) - \tilde Y(x, \pi - \theta) 
= \tilde V(x, \pi - \theta) - \tilde V(x, \theta) 
= - Q_-(x, \pi - \theta) . 
\label{anti_symm_Q}
\end{align}

\end{document}